\documentclass[a4paper,fleqn]{cas-dc}

\usepackage[authoryear,longnamesfirst]{natbib}
\usepackage{tikz}
\usepackage{orcidlink}
\usepackage{hyperref}
\usepackage{threeparttable}
\usepackage[breakable]{tcolorbox}
\usepackage{colortbl}
\usepackage[ruled,vlined]{algorithm2e}

\usepackage{booktabs}
\usepackage{enumitem}
\usepackage{amsmath,amssymb,amsfonts}
\usepackage{algorithmic}
\usepackage{makecell}
\usepackage{booktabs}
\usepackage{arydshln}
\usepackage{multirow}
\usepackage{subcaption}
\usepackage{textcomp}
\usepackage[dvipsnames]{xcolor}
\usepackage{tabularray}
 
\usepackage{array}
\usepackage{adjustbox}
\usepackage{listings}
\usepackage{xcolor}
\usepackage{caption}

\usepackage{pifont}

\usepackage[normalem]{ulem}

\definecolor{c1}{HTML}{1685a9}
\def\tsc#1{\csdef{#1}{\textsc{\lowercase{#1}}\xspace}}
\tsc{WGM}
\tsc{QE}
\colorlet{punct}{red!60!black}
\definecolor{background}{HTML}{EEEEEE}
\definecolor{delim}{RGB}{20,105,176}
\colorlet{numb}{magenta!60!black}

\lstdefinelanguage{json}{
    basicstyle=\normalfont\ttfamily,
    numbers=left,
    numberstyle=\scriptsize,
    stepnumber=1,
    numbersep=8pt,
    showstringspaces=false,
    breaklines=true,
    frame=lines,
    backgroundcolor=\color{background},
    literate=
     *{0}{{{\color{numb}0}}}{1}
      {1}{{{\color{numb}1}}}{1}
      {2}{{{\color{numb}2}}}{1}
      {3}{{{\color{numb}3}}}{1}
      {4}{{{\color{numb}4}}}{1}
      {5}{{{\color{numb}5}}}{1}
      {6}{{{\color{numb}6}}}{1}
      {7}{{{\color{numb}7}}}{1}
      {8}{{{\color{numb}8}}}{1}
      {9}{{{\color{numb}9}}}{1}
      {:}{{{\color{punct}{:}}}}{1}
      {,}{{{\color{punct}{,}}}}{1}
      {\{}{{{\color{delim}{\{}}}}{1}
      {\}}{{{\color{delim}{\}}}}}{1}
      {[}{{{\color{delim}{[}}}}{1}
      {]}{{{\color{delim}{]}}}}{1},
}



\begin{document}
\let\WriteBookmarks\relax
\def\floatpagepagefraction{1}
\def\textpagefraction{.001}

\shorttitle{}

\title [mode = title]{WebMAC: A Multi-Agent Collaborative Framework for Scenario Testing of Web Systems}

\author[aff1]{Zhenyu Wan}
\cormark[1]
\cortext[Co-first]{Co-first authors.}
\ead{wan_zy@whu.edu.cn}

\author[aff1]{Gong Chen}
\cormark[1]
\ead{chengongcg@whu.edu.cn}

\author[aff2]{Qing Huang}
\ead{qh@whu.edu.cn}

\author[aff1]{Xiaoyuan Xie}
\cormark[2]
\cortext[Corresponding]{Corresponding author.}
\ead{xxie@whu.edu.cn}

\address[aff1]{School of Computer Science, Wuhan University, China}
\address[aff2]{School of Computer and Information Engineering, Jiangxi Normal University, China}


\markboth{Journal of \LaTeX\ Class Files,~Vol.~14, No.~8, August~2021}%
{Shell \MakeLowercase{\textit{et al.}}: A Sample Article Using IEEEtran.cls for IEEE Journals}



\begin{abstract}
Scenario testing is an important technique for detecting errors in web systems. 
Testers draft test scenarios and convert them into test scripts for execution. 
Early methods relied on testers to convert test scenarios into test scripts. 
Recent LLM-based scenario testing methods can generate test scripts from natural language descriptions of test scenarios. 
However, these methods are not only limited by the incompleteness of descriptions but also overlook test adequacy criteria, making it difficult to detect potential errors. 
To address these limitations, this paper proposes WebMAC, a multi-agent collaborative framework for scenario testing of web systems. 
WebMAC can complete natural language descriptions of test scenarios through interactive clarification and transform adequate instantiated test scenarios via equivalence class partitioning. 
WebMAC consists of three multi-agent modules, responsible respectively for completing natural language descriptions of test scenarios, transforming test scenarios, and converting test scripts. 
We evaluated WebMAC on four web systems. 
Compared with the SOTA method, WebMAC improves the execution success rate of generated test scripts by 30\%–60\%, increases testing efficiency by 29\%, and reduces token consumption by 47.6\%. 
Furthermore, WebMAC can effectively detect more errors in web systems.

\end{abstract}


\maketitle

\section{Introduction}
Web systems play an important role in modern digital services. 
They support user interactions, business-process execution, and data presentation, providing a flexible and efficient operating environment for a wide range of applications \citep{amalfitano2011gui, mirshokraie2014guided, moran2017crashscope}. 

\begin{figure}
    \centering
    \includegraphics[width=0.45\textwidth]{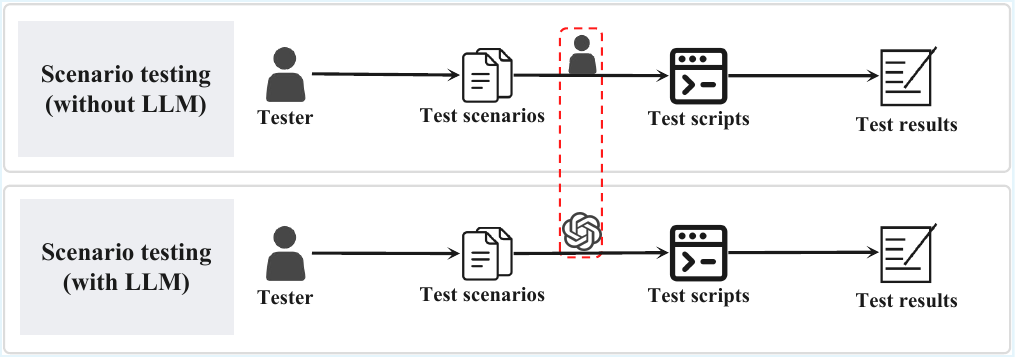}
    \caption{The technical evolution of scenario testing}
    \label{fig: scenario testing}
    \vspace{-4mm}
\end{figure}

Scenario testing is an important technique for detecting errors in web systems \citep{dalal1998factor, chandorkar2022exploratory, irshad2021adapting, zampetti2020demystifying}. 
Testers draft test scenarios and convert them into test scripts for execution.
Fig.~\ref{fig: scenario testing} shows the technical evolution of scenario testing. 
Before the advent of large language models (LLMs), early methods relied on testers to convert test scenarios into test scripts, which limited the level of automation in scenario testing.

In recent years, LLMs have demonstrated remarkable capabilities in understanding natural language and generating code. 
Researchers have introduced LLMs into scenario testing to automatically convert test scenarios into executable test scripts. 
Testers only need to design test scenarios described in natural language, and then provide these descriptions as prompts to the LLM to generate executable test scripts \citep{chen2022codet, huang2023agentcoder, shinn2023reflexion}. 
Although LLM-based scenario testing methods have further improved the automation of scenario testing, they still face \textbf{two major limitations}.

\textbf{First, incomplete natural language descriptions of test scenarios lead LLMs to generate incorrect test scripts.} 
Testers often focus on key test fields while neglecting other necessary ones. 
For example, as shown in Fig.~\ref{fig: example}(a), the add owner page requires all necessary fields to be filled before submission. 
When incomplete natural language descriptions of test scenarios are used as prompts for LLMs, missing any of these fields will lead the LLMs to generate incorrect test scripts. 
These test scripts will fail to execute due to missing required fields. 
Existing LLM-based scenario testing methods attempt to infer and supplement missing information based on the feedback from execution. 
However, these methods are inefficient, token-intensive, and often generate test scripts that do not meet the intended testing objectives \citep{bergsmann2024first, liu2023fill, li2024large, yu2025llm}.

\textbf{Secondly, in the process of converting test scenarios into corresponding test scripts, existing methods overlook test adequacy criteria (e.g., combinatorial coverage criteria).}  
For example, as shown in Figure \ref{fig: example}(b), they generate a \textbf{single} test script from an instantiated test scenario, which limits their ability to detect potential errors in the web system. 
An intuitive approach is to leverage LLMs to modify parameters and corresponding test oracles in initial instantiated test scenarios, thereby generating adequate instantiated scenarios and corresponding test scripts to detect more errors in web systems. 
However, this approach faces a key challenge: LLMs struggle to correctly understand the semantic relationship between modified parameters and test oracles, making it difficult to generate correct test scripts.

To address the two limitations mentioned above, we propose WebMAC, a multi-agent collaborative framework for scenario testing of web systems. 
WebMAC consists of three multi-agent modules: a clarification module, a transformation module, and a testing module. 
Specifically, the clarification module completes natural language descriptions of test scenarios through interactive clarification, the transformation module produces adequate instantiated test scenarios through equivalence class partitioning, and the testing module automatically generates and executes test scripts. 

\textbf{First, to address the issue of LLMs generating incorrect test scripts due to incomplete natural language descriptions of test scenarios, we design a clarification module.} 
This module completes natural language descriptions of test scenarios through collaboration among multiple intelligent agents and interactive clarification with the tester. 
Specifically, the clarification module retrieves information from the system under test (such as forms, components, etc.) and analyzes it in conjunction with the test scenario description to identify missing information. 
Then it generates clarification questions based on the identified missing information and asks the testers to clarify the missing information. 
After receiving the tester's answers, the module automatically integrates the interaction context to complete natural language descriptions of test scenarios. 
We also designed a testing module capable of converting test scenarios into executable test scripts, automatically executing them, and outputting test reports.

\textbf{Secondly, to address the issue that existing methods overlook test adequacy criteria, we added a transformation module between the clarification module and the testing module.} 
This module generates adequate instantiated test scenarios through multi-agent collaboration and equivalence class partitioning. 
Specifically, it first extracts concrete parameter values from the initial instantiated test scenario and replaces them with placeholders to construct a parameterized scenario template. 
For each extracted parameter, the module queries an external knowledge base to retrieve its corresponding equivalence class partitions. 
Then, the module automatically generates different equivalence classes based on the retrieved equivalence class partitions. 
These equivalence classes are combined and then used to instantiate the scenario template. 
The module then updates the test oracle according to the selected combinations of equivalence classes, thereby generating adequate instantiated test scenarios.

We conducted experiments on four open-source web systems to evaluate the performance of WebMAC. 
The results show that WebMAC consistently outperforms the baseline across multiple metrics. 
WebMAC can efficiently and effectively detect errors in web systems.
Specifically, WebMAC reduces the average testing time by 29\% and LLM token consumption by 47.6\%, while requiring fewer interactions.
Meanwhile, the generated test scripts achieve a higher execution success rate. 
Furthermore, by generating adequate instantiated test scenarios and converting them into test scripts, WebMAC detects more errors in web systems.
In summary, the contributions of this work are as follows:
\begin{itemize}
    \item
    We complete natural language descriptions of test scenarios through multi-agent collaboration and interactive clarification, addressing incorrect test script generation caused by incomplete descriptions of test scenarios.
    
    \item 
    We introduce combinatorial coverage criteria into the LLM-based scenario testing method, generating adequate instantiated test scenarios and corresponding test scripts through multi-agent collaboration and equivalence class partitioning.

    \item
    Experimental results show the effectiveness of WebMAC in detecting errors of web systems.
    We release WebMAC including the code and experimental data online at \cite{artifact}.
    
\end{itemize}

\section{Background and Motivation}
\begin{figure*}
    \centering
    \includegraphics[width=1\textwidth]{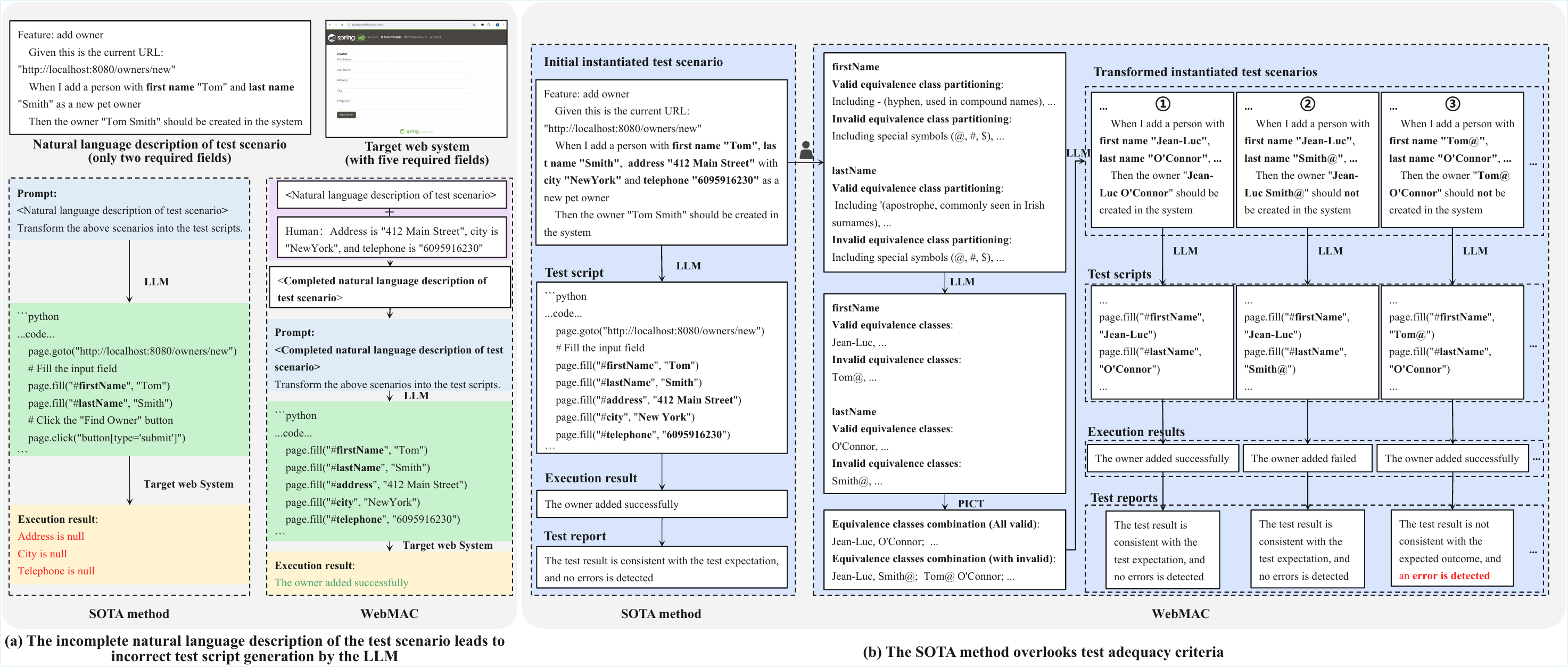}
    \caption{Motivation Example. (a) Illustrates how the incomplete natural language description of the test scenario leads to incorrect test script generation by the LLM. The natural language description of the test scenario is used as input, and the LLM converts it into a corresponding test script, which is then used to test the target web system. (b) Demonstrates how the SOTA method overlooks test adequacy criteria when converting instantiated test scenarios into test scripts. Although there are five different fields in the initial instantiated test scenario, we selected two (fast name and last name) as examples due to image size limitations.}
    \label{fig: example}
    \vspace{-4mm}
\end{figure*}
Software testing is a crucial practice to ensure the correctness, reliability, and security of software systems. 
Among various testing paradigms, scenario testing offers unique value in validating systems that involve rich user interactions or real-world operational conditions. 
Scenario testing uses scenarios that describe the interactions between users, components, or external systems in specific situations to design and execute test cases that reflect actual usage patterns.

In recent years, LLMs, represented by GPT-3.5 and its subsequent models, have demonstrated powerful capabilities, offering unprecedented opportunities to automate and enhance software engineering tasks, including software testing \citep{wang2024testeval, yuan2024evaluating}. 
In scenario testing, LLMs are regarded as natural tools for interpreting and understanding scenario descriptions provided by developers, domain experts, or requirement documents, owing to their strong natural language comprehension and generation abilities \citep{arora2024generating, wang2024software}. 
Researchers have actively explored the use of LLMs to generate test scripts from test scenarios. 
By supplying LLM with a test scenario described in natural language, the model can generate a corresponding executable test script. 
This paradigm has the potential to improve testing efficiency and reduce the effort required to convert test scenarios into executable test scripts.

A recent study explored the use of LLMs to conduct scenario testing on web systems \citep{bergsmann2024first}. 
They described test scenarios in the Gherkin format, a structured natural language used to specify business requirements in behavior-driven development. 
These scenarios were then used to construct prompts for the LLM, which generated corresponding test scripts. 
This work demonstrates that LLMs are capable of automating the conversion from test scenarios to test scripts.
However, \textbf{when faced with incomplete natural language descriptions of test scenarios, their method generates incorrect test scripts}. 
For example, as shown in Fig.~\ref{fig: example}(a), the target web system provides a form containing five fields, including \textit{last name}, \textit{first name}, \textit{city}, \textit{address}, and \textit{telephone}, while the test scenario describes an ``add owner'' case but mentions only the \textit{last name} and \textit{first name}. 
When the SOTA method uses incomplete natural language descriptions of test scenarios to construct prompts for the LLM. 
Consequently, the generated test script only requires \textit{last name} and \textit{first name} fields to be entered into the target web system. 
After executing the test script, the target web system provides an execution result indicating that the content is missing. 
Even if they use the execution result to repair the test script, it is difficult to generate a correct test script, which wastes time and tokens. 
WebMAC uses human answers to complete the description of the test scenario and then uses this description to construct a prompt, which is subsequently input into the LLM. 
This enables the LLM to correctly generate a test script with the five required fields, thereby ensuring the successful submission of the test script and improving the testing efficiency while reducing token consumption. 
Therefore, completing natural language descriptions of test scenarios before generating the test script is crucial.

Furthermore, \textbf{during the conversation of test scenarios into test scripts, existing methods overlook test adequacy criteria.} 
For example, as shown in Fig.~\ref{fig: example}(b), the SOTA method can convert instantiated test scenarios into corresponding test scripts. 
However, this test script can only cover one parameter combination (e.g., \textit{first name} ``Tom'', \textit{last name} ``Smith'', \textit{address} ``412 Main Street'', \textit{city} ``New York'', and \textit{telephone} ``6095916230''). 
It cannot cover other equivalence classes, such as empty fields, illegal characters, format errors, or boundary values. 
This makes it difficult to detect potential errors in the web system. 
WebMAC adheres to test adequacy criteria, using equivalence class partitioning to modify the parameters in the initial instantiated test scenarios to their corresponding equivalence classes. 
For example, it transforms ``Tom'' into the valid equivalence class ``Jean-Luc'', or the invalid equivalence class ``Tom@''. 
Then, it combines equivalence classes with different parameters to generate instantiated test scenarios, thus covering cases such as illegal characters. 
It also updates the test oracle to ensure the correctness of scenarios. 
For example, in scenario \ding{174} depicted in the Fig.~\ref{fig: example}(b), the scenario contains the combination ``Tom@ O'Connor'', which includes an invalid equivalence class. 
Accordingly, WebMAC modifies the test oracle to ``Then the owner ``Tom@ O'Connor'' should not be created in the system''. 
WebMAC then converts these scenarios into test scripts and executes them, detecting potential errors by comparing the execution results against the test oracle. 
In scenario \ding{174}, the execution result was ``The owner added successfully''. 
However, since ``Tom@'' constitutes an invalid equivalence class for the first name, the addition should not have succeeded. 
As the execution result did not align with the test oracle, an error was detected. 
This demonstrates that adhering to test adequacy criteria can detect potential errors.

In this paper, we crawl and filter the HTML information from the target web system to analyze the completeness of the scenario description by examining the parameters they specify. 
When the description of the test scenario is complete, no clarification is required. 
For the incomplete scenario description, we analyze missing information and pose targeted clarification questions to the tester. 
The description is then completed by incorporating the tester's answer. 
To satisfy test adequacy criteria during test script generation, we modify the parameters in the instantiated test scenario using equivalence class partitioning. 
Furthermore, we update the test oracle based on combinations of equivalence classes, thereby generating adequate instantiated test scenarios. 
Ultimately, the LLM converts them into test scripts to detect errors in web systems.

\begin{figure*}
    \centering
    \includegraphics[width=0.9\textwidth]{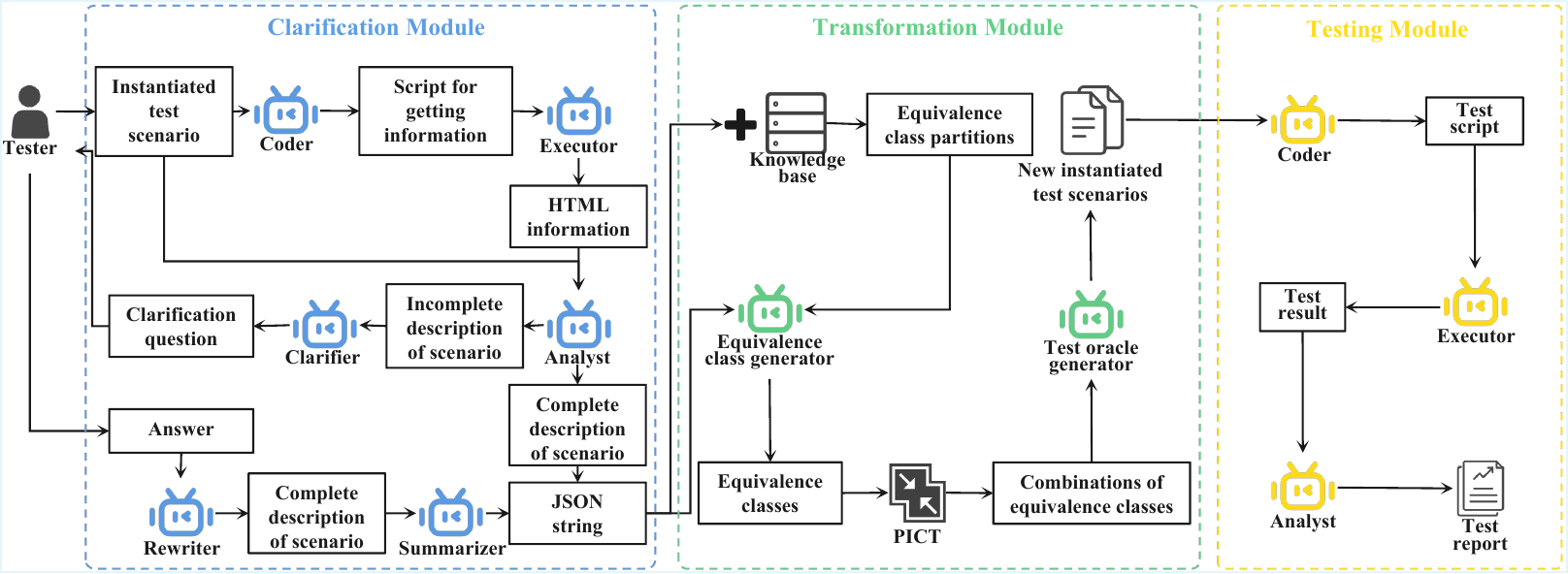}
    \vspace{-3mm}
    \caption{Overview of WebMAC}
    \label{fig: overview}
    \vspace{-3mm}
\end{figure*}

\section{Approach}
To address the limitations of incomplete natural language descriptions of test scenarios and the existing methods that overlook test sufficiency criteria, this paper proposes WebMAC, a multi-agent collaborative framework for scenario testing of web systems. 
WebMAC can complete natural language descriptions of test scenarios, generate adequate instantiated test scenarios, and convert them into corresponding test scripts. 
The approach consists of three modules: the clarification module, the transformation module, and the testing module. The overview of WebMAC is shown in Fig.~\ref{fig: overview}.

First, the tester provides an instantiated test scenario described in Gherkin format. 
In the clarification module, the Coder writes a script to crawl web forms based on the URL of the target web in the description of the test scenario, and the Executor runs the script to obtain the HTML information. 
Then, the Analyst analyzes whether the description of the test scenario is complete based on the parameters contained in the test scenario and the HTML information in the web system. 
If the description of the test scenario is complete, proceed to the next module. 
If the description of the test scenario is incomplete, the Clarifier generates clarification questions for the tester and then incorporates the tester's answers to complete the description. 
Finally, a JSON string is generated based on the historical records, which contains contextual information about the test scenario (Section~\ref{Section 2.1}). 
We use this JSON string in the transformation module. 
In the transformation module, the system retrieves the equivalence class partitions of parameters in the test scenario from an external knowledge base. 
Leveraging the retrieved partitions, the module prompts LLMs to generate different equivalence classes for the corresponding parameters. 
To ensure adequate instantiated test scenarios, we use the pairwise independent combinatorial testing tool (PICT) to combine equivalence classes of different parameters into varied combinations. 
Then, based on the combinations of equivalence classes, the test oracle is adjusted accordingly to generate adequate instantiated test scenarios (Section~\ref{Section 2.2}). 
In the testing module, once it receives an instantiated test scenario, the module generates the corresponding test script, executes it against the web system, and outputs the test report (Section~\ref{Section 2.3}).

\subsection{Clarification Module}
\label{Section 2.1}
\begin{figure}
    \centering
    \includegraphics[width=0.45\textwidth]{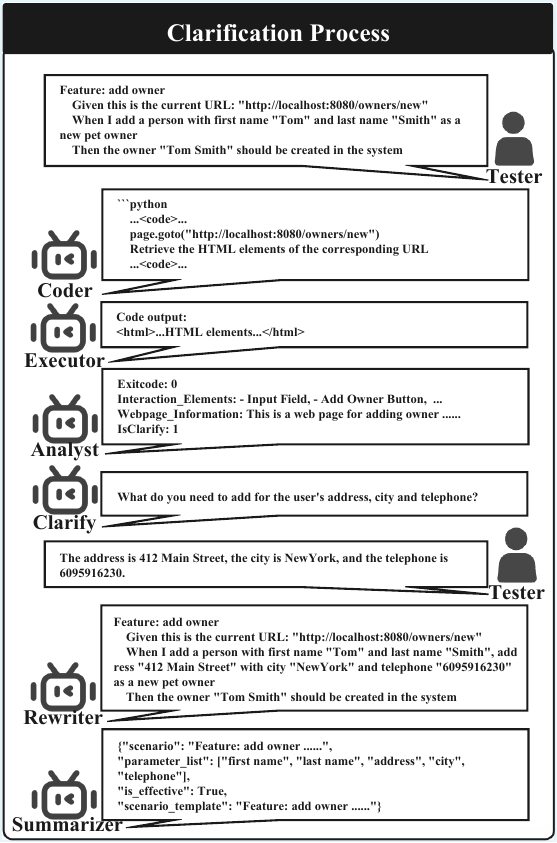}
    \caption{Clarification Process in WebMAC}
    \label{fig: clarification}
    \vspace{-2mm}
\end{figure}
The clarification module aims to improve the execution success rate of generated test scripts by leveraging human-computer interaction and multi-agent collaboration to generate fully specified test scenarios. 
In this paper, test scenarios are presented in Gherkin form. 
These scenarios contain a wealth of testing information, including the URL of the web system under test, the test behaviors, and the test oracle. 
Using these test scenarios as input prompts, the LLM can generate corresponding test scripts.

If the natural language descriptions of the test scenario is complete, then the LLM can convert the test scenario into the corresponding test script and execute it correctly. 
However, if the description of the test scenario lacks necessary form information from the target web system, the test script generated by LLMs may fail to satisfy the test oracle defined in the scenario. 
For example, consider a test scenario where a user successfully registers. 
The target webpage contains three form fields: a username field, a password field, and a telephone field. 
However, the test scenario only mentions the correct username and the correct password, missing the crucial telephone field. 
As a result, the LLM may generate a test script that leads to registration failure, violating the test oracle. 
If the correct telephone information were specified in the scenario beforehand, the test script generated by the LLM would successfully execute the registration process, thus satisfying the test oracle.

To complete the natural language description of the test scenario, the module can analyze the missing information in the test scenario by obtaining information from the web system, and then raise clarification questions to the testers regarding the missing information, thereby refining the test scenario.
The missing information in the descriptions of the test scenario typically corresponds to interactive elements on the web system. 
However, the full web system information contains substantial noise, which may hinder the LLM's ability to analyze the test scenario and generate accurate clarification questions. 
To address this issue, we use regular expressions to filter the full web system information and retain only the interactive elements, thereby reducing noise.

Specifically, we design a multi-agent system as the clarification module. 
In the clarification module, we break down the clarification process into a series of collaborative and verifiable subtasks, and introduce six agents to assist testers in clarifying test scenarios: Coder, Executor, Analyst, Clarifier, Rewriter, and Summarizer. 
These agents will collaborate to obtain information about the web system under test and help testers clarify descriptions of test scenarios. 
The functions of each agent are as follows:

\begin{itemize}
    \item
    \textbf{Coder} writes a script that can crawl the HTML information of the target web system based on the description of the test scenario.
    \item
    \textbf{Executor} automatically executes the script returned by Coder and reports the execution results.
    \item
    \textbf{Analyst} analyzes the execution results, especially the component information and form information of the target web system.
    \item
    \textbf{Clarifier} asks testers clarifying questions based on the analysis results, combined with the description of the test scenario.
    \item
    \textbf{Rewriter} completes the description of the test scenario based on the tester's answer.
    \item
    \textbf{Summarizer} summarizes the chat history and returns contextual information such as the complete description of the test scenario, parameters in the scenario, scenario template, and other information in JSON string format.
\end{itemize}

Fig.~\ref{fig: clarification} shows a specific clarification process. 
First, the tester uses the Gherkin language to write the initial instantiated test scenario and input it into the clarification module. 
The \textit{Feature} in the test scenario can be regarded as the scenario overview, the \textit{Given} statement can be regarded as the target web system to be tested, the \textit{When} statement can be regarded as the test input, and the \textit{Then} statement can be regarded as the test oracle.
According to the built-in process of the clarification module, the Coder will be called to write the script to obtain the HTML information of the target web system. 
In this example, the URL of the target web system is ``http://localhost:8080/owners/new''. 
Accordingly, the script generated by the Coder is designed to retrieve the HTML information associated with this URL. 
Once the Coder completes the script, it is passed to the Executor, which executes the script to obtain the HTML information of the web system. 
Through this collaboration, the module collects the web system information that supports subsequent clarification steps.

However, the full HTML information often contains a lot of noise, which can hinder the agents' analysis of the target web system. 
To address this issue, we apply regular expressions to filter the complete web system information and retain only the interactive elements, thereby reducing noise. 
The filtered HTML information is then passed to the Analyst for further analysis. 
The Analyst will analyze four aspects: script execution (whether it runs successfully), web system functional components, web system description, and whether the test scenario needs to be clarified. 
In Fig.~\ref{fig: clarification}, the Analyst gives its analysis results, ``Exitcode: 0'' indicates that the script is executed successfully, web system component information ``Interaction\_Elements: ...'', web system information ``Webpage\_information: This is a web page for adding owner ......'', and ``IsClarify: 1'' indicates that the scenario needs to be clarified. 
When ``IsClarify: 1'' is reported in the analysis result, the Clarifier will combine the chat history to ask clarification questions for the test scenario. 
The \textit{When} statement in the initial instantiated test scenario only contains the first name and last name fields, while the webpage contains five form components, namely \textit{first name}, \textit{last name}, \textit{address}, \textit{city}, and \textit{telephone}. 
Therefore, the Clarifier asks a clarification question for the components not mentioned in the test scenario, ``What do you need to add for the user's address, city, and telephone?''. 
This clarification question will be sent to the tester and awaits their answer. 
In response to the clarification question, the tester gives the answer ``The address is 412 Main Street, the city is NewYork, and the telephone is 6095916230.''.
Finally, the Rewriter completes the initial instantiated test scenario in combination with the answer of the tester. 
In addition, in order for the transformation module to work smoothly, we also designed an intelligent agent Summarizer to summarize the chat history of the clarification module, and summarize information such as ``scenario'', ``parameter\_list'', ``is\_effective'', and ``scenario\_template'' from the chat history. 
They are collected into a JSON string as the input of the transformation module.

\subsection{Transformation Module}
\label{Section 2.2}
\begin{figure}
    \centering
    \includegraphics[width=0.45\textwidth]{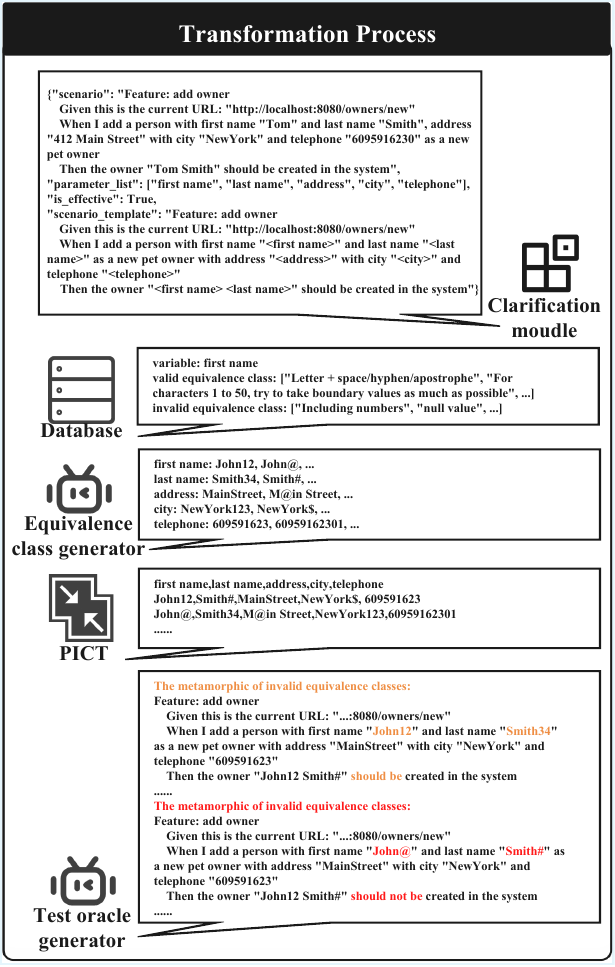}
    \caption{Transformation Process in WebMAC}
    \label{fig: metamorphic}
    \vspace{-4mm}
\end{figure}
The purpose of transforming test scenarios is to generate more adequate instantiated test scenarios based on the complete description of the test scenario output by the clarification module to detect more errors in web systems. 
In general, the parameters contained in a test scenario correspond to the form fields of the target web system under test. 
Therefore, we first analyze the JSON string produced by the clarification module. 
This JSON string includes the parameters identified in the test scenario, as well as a scenario template in which these parameters are replaced with placeholders.

For each parameter, we query an external knowledge base to retrieve the corresponding equivalence class partitions. 
The parameters and their associated equivalence class partitions are then provided to the LLM as prompts, enabling the LLM to generate a wide range of candidate equivalence classes. 
To further enhance the diversity of the generated instantiated test scenarios, we combine these equivalence classes across different parameters and fill each resulting combination into the scenario template, thereby creating adequate instantiated test scenarios.
Since a test scenario also contains test oracles, inserting a combination of equivalence classes into the scenario requires that the test oracle be updated accordingly. 
Therefore, we can adjust the test oracle accordingly based on the combination of equivalence classes, thereby achieving the transformation of instantiated test scenarios.

To effectively transform test scenarios, we design a multi-agent system as the transformation module. 
This module consists of a retrieval tool, a PICT tool, and two agents: the Equivalence Class Generator and the Test Oracle Generator. 
These agents leverage the contextual information provided by the clarification module to retrieve equivalence class partitions, generate equivalence classes, and transform test scenarios, thereby producing adequate instantiated test scenarios. 
The functions of the two agents are as follows: 

\begin{itemize}
    \item
    \textbf{Equivalence class generator} can generate equivalence classes for corresponding parameters according to the equivalence class partitions.
    \item
    \textbf{Test oracle generator} can adjust the test oracle accordingly based on the combination of equivalence classes.
\end{itemize}

Fig.~\ref{fig: metamorphic} shows a specific transformation process. 
The JSON string generated by the clarification module is provided as input to the transformation module.
Based on \textit{Feature} of the instantiated test scenario, the transformation module retrieves equivalence class partitions from an external knowledge base. 
This knowledge base stores a large collection of scenarios, parameters, and their corresponding equivalence class partitions. 
To obtain the equivalence class partitions corresponding to parameters, the module first retrieves the relevant scenario. 
For example, if \textit{Feature} of the instantiated test scenario is ``add owner'', it searches the knowledge base for scenarios associated with the keyword ``add owner''. 
It then retrieves the equivalence class partitions of the parameters listed in ``parameter\_list'' from that scenario.
For instance, the parameter \textit{first name} may have valid partitions such as ``Letter + space/hyphen/apostrophe'' and ``For characters 1 to 50, try to take boundary values as much as possible'', and invalid partitions such as ``Including numbers'' and ``null value''. 

After organizing these partitions, the module constructs a prompt that pairs each parameter with its corresponding equivalence class partitions and sends it to the Equivalence Class Generator. 
This agent then produces concrete equivalence classes according to the retrieved equivalence class partitions. 
For example, for the parameter \textit{first name}, if one partition specifies ``Including special symbols (@, \#, \$)'', the generated equivalence class may be ``John@''.
To expand the number of generated test scenarios, we use the PICT tool to create pairwise combinations of all equivalence classes, thereby producing a large set of combinations. 
However, these combinations merely modified the parameters of the test inputs within the instantiated test scenarios, whereas the content of the test oracles remained unchanged. 
To address this issue, we employ the Test oracle generator, which revises the test oracle accordingly based on the combination of equivalence classes. 
If all equivalence classes filled into the scenario template are valid, the oracle is set to a positive outcome. 
Otherwise, the oracle is modified to reflect a negative outcome. 
Through this process, the module can generate adequate instantiated test scenarios.

\subsection{Testing Module}
\label{Section 2.3}
\begin{figure}
    \centering
    \includegraphics[width=0.45\textwidth]{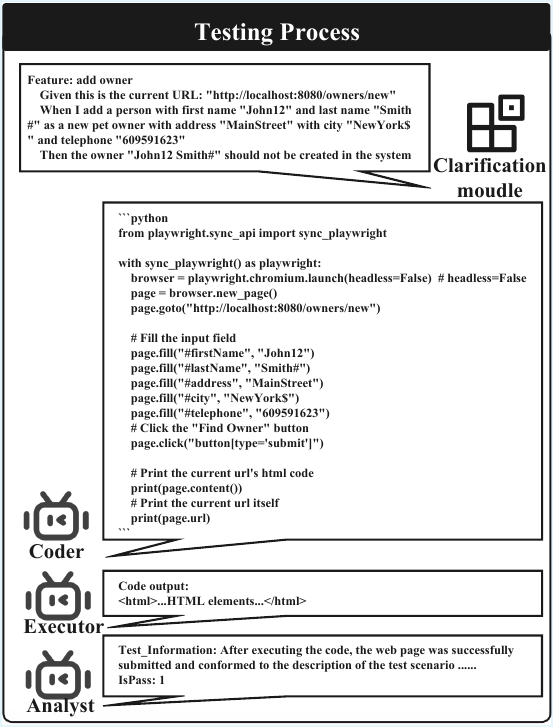}
    \caption{Testing Process in WebMAC}
    \label{fig: test}
    \vspace{-4mm}
\end{figure}
In order to convert test scenarios into corresponding test scripts and execute the test scripts, we design a multi-agent system as the testing module. 
This module consists of three agents: Coder, Executor, and Analyst, which work collaboratively to perform automated testing and return the corresponding test reports. 
The functions of these three agents are as follows: 
\begin{itemize}
    \item
    \textbf{Coder} converts the test scenario into the test script for the target web system.
    \item
    \textbf{Executor} automatically runs the test script produced by the Coder and returns the execution results.
    \item
    \textbf{Analyst} analyzes the execution results, generates both the test outcome and the corresponding test report.
\end{itemize}

Fig.~\ref{fig: test} shows a specific testing process. 
We iteratively feed the instantiated test scenarios output by the transformation module into the testing module. 
For instance, based on the descriptions contained in a test scenario: ``When I add a person with first name `John12' and last name `Smith\#' as a new pet owner with address `MainStreet' with city `NewYork\$' and telephone `609591623''', the agent Coder writes the test script to submit the owner information for the URL ``http://localhost:8080/owners/new''. 
The test script is automatically executed by the Executor and returns the execution results. 
Finally, the Analyst analyzes the execution results and reports the test execution status, thus completing the test. 
For instance, ``IsPass: 1'' means the test passed, and ``Test\_Information'' contains the test report ``After executing the code, the web page was successfully submitted and conformed to the description of the test scenario ......''.

\section{Experiments Setup}
This section presents the research design of this study, including the research questions, baseline methods, target web systems, and experimental environment.

\subsection{Research Questions}
To evaluate WebMAC, we conducted experiments to investigate the following three research questions:
\begin{itemize}
    \item
    RQ1: How does WebMAC compare to the SOTA method on test script execution success rate?
    
    \item
    RQ2: How efficient is the testing of WebMAC?

    \item
    RQ3: How effective is WebMAC in detecting errors in the web system?
    
\end{itemize}

\subsection{Baseline}
To evaluate WebMAC, we adopted the SOTA LLM-based scenario testing method proposed by Bergsmann et al. \citep{bergsmann2024first} as our baseline.
This method proposes a multi-agent system based on AutoGen that uses a Collaborator agent to schedule different agents and provide high-level guidance for the testing process.
The system takes the natural language description of the test scenario as input. 
Under the coordination of the Collaborator agent, multiple intelligent agents collaboratively analyze the scenario, generate an executable test script through iterative information exchange, and automatically execute the test script to complete the test (capable of performing actions such as clicking and inputting in the web system).

In RQ1 and RQ2, since the SOTA method does not include the capability to transform instantiated test scenarios, we compare them with a version of WebMAC where the transformation module is removed. 
This allowed us to evaluate the execution success rate of the test scripts generated by WebMAC, as well as WebMAC's testing efficiency.

\subsection{Web Systems}
We selected the Petclinic\footnote{https://github.com/spring-projects/spring-petclinic} web system in \citep{bergsmann2024first} as the system under test. 
To evaluate the applicability of our approach across different web systems, we used keywords such as ``Java web'' and ``HTML'' to search for web systems on GitHub and cloned 14 projects. 
After excluding web systems that could not run properly in our experimental environment, we finally selected three additional web systems with different functionalities (Blog\footnote{https://github.com/mangeshpawar830/spring-boot-blog-master}, Tour-reservation\footnote{https://github.com/muhammadumarrasheed/tour-reservation}, and Tracw\footnote{https://github.com/NithinBairoju/tracw}) for our experiments.

PetClinic is a pet hospital system that allows users to add, search for, and access pets and their owners. 
Blog is a system for creating and browsing blogs. 
Tour-reservation is a travel ticketing system. 
Tracw is a wallet system that allows users to set up deposit and withdrawal limits.

\subsection{Experiment Environment}
All experiments were conducted on an HP workstation (Precision 3660) equipped with a 12th Gen Intel(R) Core(TM) i7-12700 processor and 16GB of RAM. 
The Google Chrome version used was 140.0.7339.128 (Official Build) (arm64), and the Chrome web driver version was 140.0.7339.185 (r1496484).

\section{Experiments Results}

\subsection{How does WebMAC compare to the SOTA method on test script execution success rate?}

\begin{figure}
    \centering
    \includegraphics[width=0.45\textwidth]{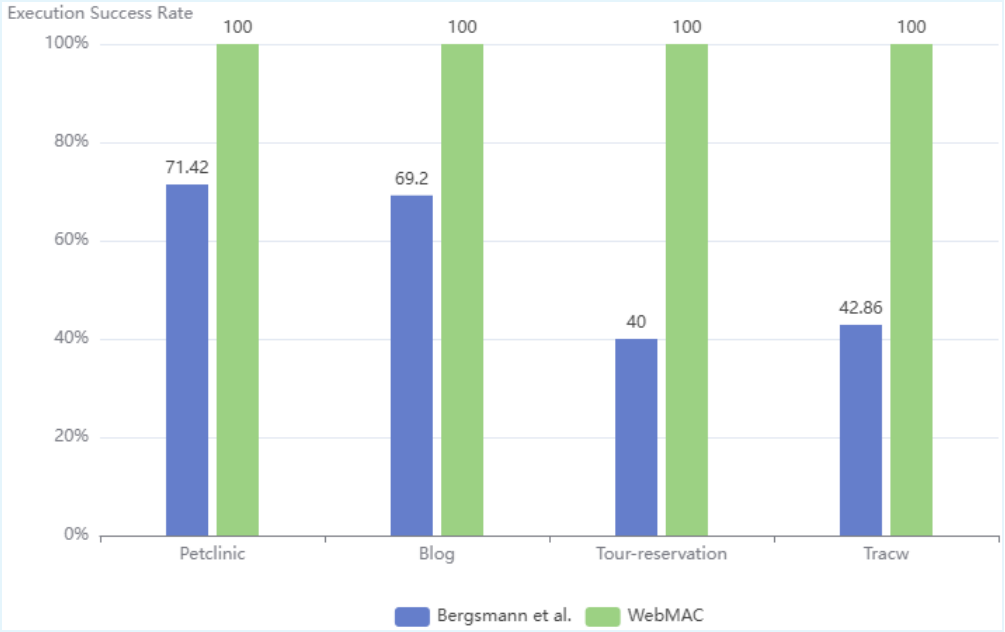}
    \caption{The execution success rate of test scripts generated by WebMAC and the SOTA method on four different web systems}
    \label{fig: RQ1}
\end{figure}

The execution success rate of the test scripts generated by WebMAC and the SOTA method on four different web systems is presented in the form of a bar chart in Fig.~\ref{fig: RQ1}. 
It can be seen that the execution success rate of the test scripts generated by our method on the four different web systems is stronger than that of the SOTA method.  
Specifically, both WebMAC and the SOTA method perform well when descriptions of test scenarios are complete, but WebMAC performs significantly better than the SOTA method when handling incomplete descriptions of test scenarios.

When handling incomplete descriptions of test scenarios, the SOTA method often generates incorrect test scripts, which leads to test script submission failures. 
Although the SOTA method can regenerate test scripts by incorporating system feedback, it typically requires multiple iterations before a correct script is generated. 
As the number and complexity of form fields in the web system increase, the number of iterations required for incomplete natural language descriptions of test scenarios also grows. 
These repeated iterations accumulate extensive interaction history filled with noise, making it more likely for the LLM to generate test scripts that deviate from the tester's intent, ultimately causing test failures.
For example, consider the ``\textit{add owner}'' test scenario for the web system Petclinic: ``\textit{Feature: Add owner; Given this is the current URL: http://localhost:8080/owners/new; When I add a person with first name `John' and last name `Smith' as a new pet owner; Then the owner `John Smith' should be created in the system}''. 
The scenario lacks concrete values for \textit{address}, \textit{city}, and \textit{telephone} fields, causing the LLM to generate a test script based solely on the incomplete natural language descriptions of test scenarios and then rely on the web system's feedback to infer which fields are missing. 
However, the web system's feedback typically provides only general error messages, such as ``\textit{address is null}'', without offering any further guidance on the missing content.
This often leads the LLM to generate arbitrary placeholder values like ``\textit{\textless address \textgreater}''. 
In contrast, WebMAC introduces human-computer interaction to provide concrete values at critical moments. 
WebMAC identifies the missing contents in the description of the test scenario and asks clarification questions to the tester. 
The tester's responses supply the specific missing content, and the LLM then uses these responses to complete the description of the test scenario. 
This process also avoids the accumulation of noisy history during iteration, which could affect the completion of the scenario description.

Additionally, we found that the SOTA method performs particularly poorly when the test scenario is negative. 
When the input test scenario describes a case in which submitting invalid input should cause the web system to reject the submission, the SOTA method still initiates automatic iterative refinement. 
It repeatedly adjusts the test script based on the web system's feedback and may even insert values into the test script that contradict the original test scenario. 
This is because the workflow designed by the SOTA method focuses on accomplishing actions rather than executing the scenario itself, which weakens the role of test oracles embedded in test scenarios. 
For example, consider the ``\textit{register account}'' test scenario for the web system Tracw, ``\textit{Feature: Register account; Given this is the current URL: http://localhost:8080/register; When I entered the Username `abc', the Password `123456@Mm', Telephone `123456789', Initial Balance `10', and then clicked `Register'; Then register failed}''. 
In this test scenario, the test should fail because the input telephone is invalid. 
However, the SOTA method instructs the LLM to achieve the goal of completing account registration. 
As a result, after receiving feedback from the webpage indicating that the telephone is invalid, the LLM proactively modifies the illegal telephone into a valid one to ensure successful registration. 
This violates the test oracle specified in the test scenario.

\vspace{1mm}
\noindent\fbox{
\begin{minipage}{7.8cm} 
\textit{In four different web systems, the execution success rate of the test scripts generated by WebMAC is \textbf{significantly higher} than the SOTA method, which proves that WebMAC can effectively improve the execution success rate of the generated test scripts.
}
\end{minipage}}

\subsection{How efficient is the testing of WebMAC?}

\begin{table*}[h]
\centering
\caption{The effectiveness and efficiency of WebMAC and the SOTA method on four different web systems}
\label{RQ2}
\vspace{-1mm}
\renewcommand{\arraystretch}{1.5}
\resizebox{0.95\textwidth}{!}{
\begin{tabular}{|c|c|c|c|c|c|c|c|c|}
\hline
Web System & Approach & Execution Success Rate & \begin{tabular}[c]{@{}c@{}}Avg. Clar.\\Time (s)\end{tabular} & \begin{tabular}[c]{@{}c@{}}Avg. Test\\Time (s)\end{tabular} & \begin{tabular}[c]{@{}c@{}}Avg. Clar.\\Tokens\end{tabular} & \begin{tabular}[c]{@{}c@{}}Avg. Test\\Tokens\end{tabular} & \begin{tabular}[c]{@{}c@{}}Avg. Interactions\\During Clarification\end{tabular} & \begin{tabular}[c]{@{}c@{}}Avg. Interactions\\During Test\end{tabular} \\ \hline
\multirow{2}{*}{Petclinic} & Bergsmann et al. & 71.43\% & 0 & 42.45 & 0 & 19867.94 & 0 & 14.27 \\ \cline{2-9} 
 & WebMAC & \textbf{100\%} & \textbf{24.26} & \textbf{12.71} & \textbf{7261.82} & \textbf{4592.17} & \textbf{7.23} & \textbf{4} \\ \hline
\multirow{2}{*}{Blog} & Bergsmann et al. & 69.20\% & 0 & 43.05 & 0 & 19717.08 & 0 & 14.85 \\ \cline{2-9} 
 & WebMAC & \textbf{100\%} & \textbf{23.72} & \textbf{8.44} & \textbf{4677.92} & \textbf{4590.38} & \textbf{7.75} & \textbf{4} \\ \hline
\multirow{2}{*}{\begin{tabular}[c]{@{}c@{}}Tour-\\reservation\end{tabular}} & Bergsmann et al. & 40\% & 0 & 36.78 & 0 & 18757.67 & 0 & 13.47 \\ \cline{2-9} 
 & WebMAC & \textbf{100\%} & \textbf{17.48} & \textbf{8.69} & \textbf{5038.07} & \textbf{5292.53} & \textbf{6.4} & \textbf{4} \\ \hline
\multirow{2}{*}{Tracw} & Bergsmann et al. & 42.86\% & 0 & 52.41 & 0 & 25473.48 & 0 & 13.38 \\ \cline{2-9} 
 & WebMAC & \textbf{100\%} & \textbf{20.60} & \textbf{8.17} & \textbf{6672.19} & \textbf{5781.71} & \textbf{7.14} & \textbf{4} \\ \hline
\end{tabular}}
\vspace{-1mm}
\end{table*}

As shown in Table~\ref{RQ2}, WebMAC achieves shorter testing time than the SOTA method across four different web systems. 
In terms of token consumption, WebMAC also uses significantly fewer tokens. 
At the same time, WebMAC requires fewer interactions with the LLM compared to the SOTA method.

In terms of testing time, the clarified test scenarios already have all missing information filled in, allowing the system to generate correct test scripts without iterative refinement.  
This enables the testing process to complete much more quickly.
Comparing the performance of the two approaches across different web systems, WebMAC reduces the average test time by 75\% compared to the SOTA method.
In Petclinic, because the web system contains relatively few form fields, the test scenarios involve fewer parameters. 
This allows the SOTA method to complete automated testing more quickly. 
WebMAC reduces the average test time by 70\% compared to the SOTA method. 
In contrast, in Tracw, the web system contains many more form fields, and the corresponding scenarios involve more parameters. 
As a result, the SOTA method requires significantly more time to complete automated testing for these descriptions of test scenarios, causing WebMAC to reduce the average test time by 84\% compared to the SOTA method. 
This indicates that the more complex the forms of the web system are, the greater the advantage of WebMAC becomes.
Of course, WebMAC requires a clarification process. 
However, even after including the time spent on clarification, WebMAC still reduces the average testing time by 29\% compared to the SOTA method.

In terms of token consumption, once the test scenario has been clarified, all missing information has already been completed. 
Therefore, no iterative repair is needed, preventing excessive token usage. 
Across different web systems, WebMAC reduces the average number of test tokens by 76\% compared to the SOTA method.
In Tour-reservation, because the web system is simple and contains less HTML information, the Executor produces fewer tokens. 
As a result, the SOTA method consumes fewer tokens in this web system. 
WebMAC reduces the average number of test tokens by 71\% compared to the SOTA method. 
In contrast, in Tracw, the web system is more complex and contains a larger amount of HTML information, leading the Executor to output more tokens. 
This further amplifies the advantage of WebMAC in token consumption, causing WebMAC to reduce the average number of test tokens by 77\% compared to the SOTA method.
Similarly, WebMAC also consumes some tokens during the clarification process. 
However, even after including the tokens used in clarification, WebMAC still consumes 47.6\% fewer test tokens compared to the SOTA method.

In terms of interaction count, WebMAC, because it uses test scenarios with already completed content, can directly generate correct test scripts without a large number of interactions. 
In contrast, the SOTA method requires iterative generation of test scripts and is affected by noise in the historical records generated during iteration, requiring more interactions to generate correct test scripts.
In terms of performance across different web systems, the SOTA method requires approximately 3.5 times more user interactions on average than WebMAC. 
When generating test scripts, the SOTA method must first produce an initial test script and then refine it based on the feedback of the web system. 
Depending on the complexity of the test scenario, the SOTA method may require two to three rounds of iteration before obtaining a correct test script.
In contrast, WebMAC first clarifies the incomplete natural language description of the test scenario, allowing WebMAC to generate the test script without any iteration, resulting in a stable interaction count of four.
Although WebMAC performs completeness checking and clarification, even after including the interactions involved in the clarification stage, WebMAC still maintains a clear advantage over the SOTA method in terms of interaction count.

\vspace{1mm}
\noindent\fbox{
\begin{minipage}{7.8cm} 
\textit{WebMAC \textbf{outperforms} the SOTA method in testing time, token consumption, and interaction count with the LLM across four different web systems, demonstrating that WebMAC can effectively reduce testing costs and improve testing efficiency.
}
\end{minipage}}

\subsection{How effective is WebMAC in detecting errors in the web system?}

\begin{table}[ht]
\centering
\caption{The errors detected in different web systems}
\label{RQ3}
\renewcommand{\arraystretch}{1.5} 
\resizebox{0.5\textwidth}{!}{
\begin{tabular}{|c|c|c|c|c|}
\hline
\textbf{Web System} & 
\textbf{Approach} & 
\textbf{\begin{tabular}[c]{@{}c@{}}The number of \\ transformation \\ scenarios\end{tabular}} & 
\textbf{\begin{tabular}[c]{@{}c@{}}The number of \\ discovered \\ errors\end{tabular}} & 
\textbf{\begin{tabular}[c]{@{}c@{}}Types of \\ errors \\ discovered\end{tabular}} \\ \hline

\multirow{2}{*}{Petclinic}        & Bergsmann et al. & 1   & 0  & 0  \\ \cline{2-5} 
                                  & WebMAC              & \textbf{119} & \textbf{26} & \textbf{14} \\ \hline

\multirow{2}{*}{Blog}             & Bergsmann et al. & 1   & 0  & 0  \\ \cline{2-5} 
                                  & WebMAC              & \textbf{93} & \textbf{21} & \textbf{13} \\ \hline

\multirow{2}{*}{\begin{tabular}[c]{@{}c@{}}Tour-\\reservation\end{tabular}} & Bergsmann et al. & 1   & 0  & 0  \\ \cline{2-5} 
                                  & WebMAC              & \textbf{84}  & \textbf{19} & \textbf{11} \\ \hline

\multirow{2}{*}{Tracw}            & Bergsmann et al. & 1   & 0  & 0  \\ \cline{2-5} 
                                  & WebMAC              & \textbf{78} & \textbf{16} & \textbf{11} \\ \hline

\end{tabular}}
\end{table}

In this experiment, we evaluate the ability of WebMAC to detect errors across four different web systems. 
As shown in Table~\ref{RQ3}, starting from an incomplete instantiated test scenario, WebMAC can clarify and transform it into hundreds of distinct instantiated test scenarios. 
These transformed scenarios are able to uncover multiple errors, and the types of errors detected vary across systems. 
This demonstrates that our approach can effectively detect errors in the web systems. 
In contrast, Bergsmann et al.'s method overlooks test adequacy criteria and generates only a single test script based on the initial test scenario, which fails to effectively expand the test coverage and therefore fails to detect errors in all four systems.

Specifically, in Petclinic, after the description of the test scenario is clarified and transformed by WebMAC, a total of 119 different instantiated test scenarios are generated. 
For example, in a test scenario describing adding an owner successfully, the form of the target web system contains \textit{first name}, \textit{last name}, \textit{address}, \textit{city}, and \textit{telephone}. 
A single instantiated test scenario can only detect the situation where all of \textit{first name}, \textit{last name}, \textit{address}, \textit{city}, and \textit{telephone} are valid. 
After transformation, however, the 119 test scenarios include cases such as: (1) an invalid \textit{first name} with valid \textit{last name}, \textit{address}, \textit{city}, and \textit{telephone}; and (2) an invalid \textit{last name} with valid \textit{first name}, \textit{address}, \textit{city}, and \textit{telephone}, among others. 
These test scenarios cover all the equivalence class partitions in the external knowledge base, thereby revealing 26 errors, including 14 distinct types of errors.
For instance, for the invalid equivalence class ``\textit{the first name must not contain @}'', the corresponding equivalence-class value ``\textit{John@}'' surprisingly passes the test, revealing that the target web system ignores the illegal-character restriction on \textit{first name}. 
Another example is the valid equivalence class for \textit{telephone}, ``\textit{area codes may be connected with hyphens}'', where the valid value ``\textit{609-591-6230}'' fails to successfully add an owner and produces an error report, revealing that the target web system does not support this correct telephone format. 
Experiments on Blog, Tour-reservation, and Tracw similarly generate adequate instantiated test scenarios through this transformation process, covering various possible cases of the target web systems, thereby detecting errors in web systems.

\vspace{1mm}
\noindent\fbox{
\begin{minipage}{7.8cm} 
\textit{WebMAC was evaluated on four different web systems, where it was able to generate adequate instantiated test scenarios and detect errors in the target web systems, demonstrating the effectiveness of WebMAC in detecting errors.
}
\end{minipage}}

\section{Discussion}
There are potential threats to the validity of our approach. 
Additionally, we discussed the impact of combining equivalence classes on scenario coverage.

\subsection{Threats to Internal Validity}
A potential threat to internal validity stems from WebMAC's reliance on OpenAI's third-party API services for test scenario clarification, transformation, and script generation. 
Network latency or instability during API calls could introduce variability in execution time or even cause request failures, thereby affecting the reproducibility and consistency of the experimental results. 
To mitigate this issue, we selected higher-speed network equipment and conducted experiments during periods of stable network conditions. 
To minimize this confounding factor, all experiments were conducted in a controlled network environment with dedicated bandwidth and during off-peak hours to ensure stable connectivity. 
Nevertheless, residual variability due to external API dependencies remains a limitation. 
Another threat to internal validity lies in the assumption that test oracles can be systematically derived from equivalence class partitions. 
In practice, for a small subset of test scenarios, especially those involving complex state transitions or dependencies on external services, it is difficult to derive the expected behavior based on equivalence classes. 
This may lead to incorrect oracles, potentially compromising the execution success rate of the test script.

\subsection{Threats to External Validity}
With respect to external validity, some target web systems may trigger behaviors that lead to non-HTML pages, from which relevant page information cannot be directly obtained. 
This limits the applicability of our approach in such scenarios. 
However, our approach can be readily adapted to such systems by updating the example template used by the Coder. 
Moreover, the equivalence class partitions are manually defined for each web system, which may hinder the portability of our approach across diverse application domains. 
Future work should investigate automated methods for deriving equivalence class partitions, thereby reducing manual effort and improving cross-domain adaptability.

\subsection{Impact of Combining Equivalence Classes}
Our approach generates diverse test scripts by combining equivalence classes. 
Effective combination of equivalence classes is crucial to improve both testing efficiency and error detection effectiveness. 
In this work, we employ the classic combinatorial testing tool PICT to generate combinations of equivalence classes. 
However, while PICT produces a large number of combinations of equivalence classes, not all instantiated test scenarios corresponding to combinations can trigger errors, which affects the efficiency of error detection to some extent. 
Consequently, a promising direction for future research is to implement a rigorous filtering process for these combinations of equivalence classes to mitigate this issue.

\section{Related Work}
This section introduces some related work, including LLM-based scenario testing, the multi-agent collaborative framework, and equivalence class partitioning.

\subsection{LLM-based Scenario Testing}
The application of LLMs in software testing has seen significant advancements in recent years, particularly in scenario testing \citep{liu2023fill, li2024large, yu2025llm}. 
Current research primarily uses the natural language understanding and code generation capabilities of LLMs to translate test scenarios into test scripts directly \citep{li2023mobile, yang2024evaluation, korraprolu2025test}. 
For example, TestGPT-Server \citep{wang2025testgpt} employs an LLM-enhanced interface analyzer and code analyzer to extract API syntax and semantics, and then generates diverse test requests through a targeted request generator and a guided random tester. 
An LLM-assisted response analyzer further identifies service crashes and detects errors through automated assertion mining. 
In GUI testing, GUIPilot \citep{liu2025guipilot} leverages LLMs for automated exploration, but its test oracle relies on manually provided mockup designs, hindering the automatic construction of accurate verification conditions. 
Similarly, GPTDroid \citep{liu2024make} formulates mobile GUI testing as a question-answering task, using LLMs to generate test scripts. 
Although these approaches improve the level of automation in scenario testing, they often lead LLMs to generate incorrect test scripts when confronted with incomplete natural language descriptions of test scenarios. 
The WebMAC framework proposed in this paper mitigates this limitation by using a multi-agent collaborative framework to assist testers in completing natural language descriptions of test scenarios and enabling the generation of correct test scripts \citep{huang2023let}.

\subsection{Multi-Agent Collaborative Framework}
As research on LLMs deepens, single agents are no longer sufficient for many tasks. 
Consequently, research has shifted to the multi-agent collaborative framework \citep{li2023camel, wu2024autogen}. 
These frameworks are being used in various research fields and have demonstrated excellent performance \citep{guo2024large, qian2024scaling, yan2025beyond}. 
Through task decomposition and role division, multi-agent systems enable different agents to focus on specific subtasks \citep{bo2024reflective, zhu2024survey}. 
Agents can share knowledge and context through interaction and collaboration, imbuing the system with collective intelligence, achieving greater efficiency and robustness in complex tasks. 
In scenario testing, multi-agent collaborative frameworks coordinate complex tasks and improve testing efficiency through division of labor and collaboration.
For example, Bergsmann et al. \citep{bergsmann2024first} use a Collaborator agent that coordinates multiple agents and provides high-level guidance for generating and executing Gherkin test scenarios. 
However, scenario testing often involves ambiguous or incomplete inputs. 
In such cases, relying on a centralized agent to orchestrate the entire workflow is prone to errors, as it may misinterpret these ambiguities and generate incorrect test scripts. 
To address this limitation, WebMAC proposes a multi-agent collaboration framework for scenario testing. 
Rather than relying solely on agents to interpret incomplete natural language descriptions of test scenarios, WebMAC integrates human-computer interaction to resolve ambiguities, thereby preventing erroneous assumptions that could lead to invalid test scripts. 
This significantly enhances the efficiency of scenario testing and the execution success rate of the generated test scripts.

\subsection{Equivalence Class Partitioning}
Equivalence class partitioning divides the input domain into disjoint subsets that are behaviorally equivalent to the system under test \citep{bhat2015equivalence, hubner2019experimental}. 
By selecting a representative input from each subset, testers can avoid redundant test cases while maintaining coverage, thereby improving testing efficiency. 
In practice, equivalence class partitioning is commonly combined with boundary value analysis, valid/invalid input classification, and combinatorial interaction coverage strategies to improve error detection effectiveness under limited testing budgets \citep{radcliffe1991equivalence, fields1987structure, fields1993effects}. 
WebMAC's transformation module extracts parameters from instantiated test scenarios and automatically derives equivalence classes for each parameter. 
WebMAC then combines the equivalence classes of different parameters using PICT \citep{pict} to generate a large number of combinations of equivalence classes. 
Finally, for each combination of equivalence classes, WebMAC synthesizes an appropriate test oracle and generates a corresponding instantiated test scenario.

\section{Conclusion}
In this paper, we propose WebMAC, a multi-agent collaborative framework for scenario testing of web systems. 
Through interactive clarification and multi-agent collaboration, WebMAC mitigates the problem of generating incorrect test scripts due to incomplete natural language descriptions of test scenarios in LLM-based scenario testing. 
Furthermore, it adheres to test adequacy criteria, leveraging multi-agent collaboration and equivalence class partitioning to correctly generate adequate instantiated test scenarios and corresponding test scripts, thereby detecting more errors in web systems. 
Experiments across multiple web systems demonstrate that WebMAC significantly improves both the execution success rate and the efficiency of test script generation, while effectively detecting errors. 
This work advances automated scenario testing and promotes deeper integration of LLMs into software engineering practices. 
Future work will further study the scale issues that may arise in the process of equivalence class combination, explore more efficient filtering and selection strategies, and extend WebMAC to more complex and diverse software test scenarios.

\section{Declaration of competing interest}
The authors declare that they have no known competing financial interests or personal relationships that could have appeared to influence the work reported in this paper.

\section{Acknowledgments}
This work was partially supported by the National Key R\&D Plan of China (Grant No.2024YFF0908003), and the Natural Science Foundation of China (No.62472326).


%

\bibliographystyle{cas-model2-names}

\bibliography{sample-base}




\end{document}